\documentstyle[12pt]{article}
\begin{document}
\thispagestyle{empty}

\title{ Dirac Hartree-Fock for Finite Nuclei\\ Employing Realistic
Forces}

\author{ R.~Fritz , H.~M\"uther \\
{\it Institut f\"ur Theoretische Physik der Universit\"at
 T\"ubingen}\\
{\it D-7400 T\"ubingen, Federal Republic of Germany}
\\ \\ and \\ \\
R.~Machleidt\\{\it Department of Physics, University of Idaho}\\
{\it Moscow, Idaho 83843, U.S.A.}}

\maketitle

\date{\today}
\begin{abstract}
We discuss
two different approximation schemes for the
self-consistent solution of the {\it relativistic} Brueckner-Hartree-Fock
equation for finite nuclei. In the first scheme, the Dirac effects are
deduced from corresponding nuclear matter calculations, whereas in the second
approach the local-density approximation is used to account for the
effects of correlations. The results obtained by the two methods are
very similar. Employing a realistic one-boson-exchange potential (Bonn~A), the
predictions for energies and radii of $^{16}$O and $^{40}$Ca come out in
substantially better agreement with experiment as compared
to non-relativistic approaches.
As a by-product of our study, it turns out that the Fock exchange-terms,
ignored in a previous investigation, are not negligible.
\end{abstract}

\clearpage
One of the most fundamental challenges of nuclear
 many-body theory is to derive the
bulk properties of nuclei, like energies and radii,
from a realistic nucleon-nucleon (NN) interaction. Here,
the term ``realistic'' refers to nuclear potentials
which reproduce the two-nucleon data accurately.
Representative examples
are the Paris NN potential~\cite{paris} and the models developed by the
Bonn group~\cite{rup}. Typically, these forces contain strong
components of short range, which make it inevitable to carefully account
for the two-nucleon correlations at short inter-nucleonic distances.
In the Brueckner-Hartree-Fock (BHF) method this is done by
solving the Bethe-Goldstone equation. The resulting G-matrix can be
understood as an effective interaction which incorporates effects of NN
correlations and depends on the properties of the nuclear system under
investigation.

For many years, calculations of this kind have been performed
with only limited success. More than 20 years ago, Coester
{\it et al.}~\cite{coester} observed that, in an energy {\it versus}
density plot, the saturation points of nuclear matter
as obtained in BHF calculations employing different realistic
potentials are located along a band (`Coester band') that does not
include the empirical value.
For example,
one may find a realistic NN
interaction which reproduces the binding energy of nuclear matter
correctly, but
at a saturation density which is about twice the empirical one.
{\it Vice versa}, a different
realistic interaction may predict the saturation density correctly
but yield a binding energy of about 11 MeV per nucleon rather than
the empirical 16 MeV.

In the past, there have been many attempts to improve nuclear
many-body theory within the framework of non-relativistic quantum
mechanics. Three-nucleon
correlations and other corrections to BHF have been considered,
however, without substantial success~\cite{day}.
A phenomenon similar to the Coester band for nuclear matter, has
been found for finite nuclei~\cite{zabo,repor,carlo}:
BHF calculations yield either correct binding energies but too small
radii, or correct radii but too small binding energies.
We mention
that Kuo {\it et al.}~\cite{tom} were able to improve the predictions
for nuclear matter by including effects of so-called
particle-particle/hole-hole ring diagrams (long-range correlations).
However, it appears that this method does not improve
the predictions for finite nuclei, sufficiently~\cite{harry}.

Motivated by the success of the $\sigma$ - $\omega$ model of Walecka
and Serot~\cite{serot}, attempts have been made to incorporate the relativistic
features of this approach also in nuclear structure calculations
which are based upon the
realistic NN force. Such Dirac BHF (DBHF) calculations,
as this has become known,
have been performed for nuclear matter by, e.~g., Shakin and
collaborators~\cite{shakin}, Brockmann and Machleidt~\cite{BM84},
 and ter Haar and Malfliet~\cite{malf}. The basic aspects of this approach
have been thoroughly investigated by Horowitz and Serot~\cite{HS84}. In
the DBHF approach, one accounts for the fact that the relativistic
nucleon self-energy in nuclear matter is given essentially
 by a large attractive
scalar (`$\sigma$') and repulsive vector (`$\omega$') field.
The single-particle motion is described by a Dirac equation which
includes this self-energy. Due to the scalar field, the
nucleon mass is reduced enhancing the ratio between small
and large components of the Dirac spinors. Moreover,
the sigma field  decouples
causing a strongly density-dependent repulsive effect.

Due to these features, Brockmann and
Machleidt~\cite{BM84,brma} were able to reproduce nuclear matter
saturation correctly in a DBHF calculation employing
a realistic one-boson-exchange
NN potential (`Bonn~A').
Now, the crucial question is
whether the
DBHF approach can also explain the
bulk properties of finite nuclei.

The self-consistent solution of the DBHF equation for finite nuclei
is much more involved than for nuclear matter. Therefore, two
different approximation schemes have been developed. In the
first scheme~\cite{mmb}, the pair correlations are
calculated in the finite nucleus under consideration
whereas the medium dependence of the Dirac
spinors is taken into account via the local-density approximation.
Thus, the Bethe-Goldstone equation is solved directly for the
finite nucleus satisfying the self-consistency requirement of
conventional BHF.
Relativistic medium effects are taken into account by evaluating
the potential matrix
elements and the kinetic energy in terms of in-medium Dirac
spinors.

In Table~1 (column DBHF), we show results of relativistic BHF
calculations performed according to scheme one for the nuclei $^{16}$O and
$^{40}$Ca. In all calculations of this note, the Bonn A potential~\cite{rup}
is applied.
It is interesting to compare these results with non-relativistic
BHF calculations, in which the medium-dependence
of the Dirac spinors is ingnored (column BHF of Table~1).
It is seen that the Dirac effects
increase the binding energy {\it and} the charge radius.
Thus, the DBHF results are in better agreement with experiment.
Typically, the remaining discrepancy between the DBHF results and
experiment is only about one
half of the corresponding discrepancy in conventional
BHF calculations.
For $^{16}$O,
results have been reported in Ref.~\cite{mmb}. They are
confirmed by our present investigations, which also includes
$^{40}$Ca.

In the second approximation scheme,
one treats the pair correlations in local-density
approximation, while the Dirac equation is solved directly for the finite
nucleus. This can be done by defining an effective medium-dependent
meson-exchange
interaction based upon the nuclear matter G-matrix.
Since the G-matrix is density-dependent, so are the coupling constants
of the effective interaction which are adjusted such as to
reproduce the G-matrix~\cite{marc,elsen}.
A simple and successful calculation
along this line has recently been reported by Brockmann
and Toki~\cite{bt}. For the effective interaction,
they consider $\sigma$ and $\omega$ exchange,
adjusting the coupling constants such that
a simple Dirac-Hartree calculation reproduces the
nuclear matter DBHF results.
The density-dependence of the resulting coupling constants is
displayed in Fig.~1. The coupling constants for both $\sigma$ and
$\omega$ decrease with increasing density. This is clearly the
correlation effect~\cite{elsen}.

Employing these density-dependent coupling constants in a Dirac-Hartree
calculation for finite nuclei, one obtains good agreement
between theory and experiment (see Ref.~\cite{bt} and column RDH in
our Table 1), keeping in mind that these are
parameter-free calculations based upon a
realistic NN interaction. Our RDH results,
displayed in Table 1, deviate slightly from those obtained by Brockmann
and Toki~\cite{bt}. The differences can be understood as follows:
First, we include a center-of-mass correction
(as discussed, e.~g., in Ref.~\cite{mmb}) to allow for a comparison
between the various approximations displayed in Table 1.
Secondly, we observed a sensitivity
of the results on the extrapolation of the coupling constants to
densities below the lowest density considered by Brockmann and Toki.
We tried to remove this sensitivity by inspection of nuclear matter at
low densities. The Dirac-Hartree, as well as the Dirac-Hartree-Fock
equation discussed below, were solved by a matrix diagonalisation
method, expanding the wavefunctions in a basis of states for a
spherical cavity~\cite{fri932}. The new computer code has been tested by
comparing the results with those presented by Bouyssy {\it et al.}~\cite{bou}.

Inspection of Table~1 reveals that there are significant differences
between the RDH and DBHF results.
Since both
schemes are approximations to a complete
self-consistent calculation for a finite nucleus, at least one
approach must be a poor approximation.

The natural step beyond the relativistic density-dependent Hartree
(RDH) approximation is
relativistic Hartree-Fock. Assuming a $\sigma$ - $\omega$ model
with density dependent coupling constants $g_{\sigma}$ and
$g_{\omega}$, the scalar part of the nucleon self-energy in nuclear
matter at density $\rho$ can be written as~\cite{serot,elsen}
\begin{eqnarray}
U^s(k,\rho ) & = & \frac{-4}{(2 \pi )^3}
\frac{g_{\sigma}^2 (\rho )}{m_{\sigma}^2} \int_{0}^{k_{F}} d^3q
\frac{M^*(q,\rho )}{E^*(q,\rho )} \nonumber \\
&&+ \frac{1}{4\pi^2 k }\int_{0}^{k_{F}} q\, dq \frac{M^*(q,\rho )}
{E^*(q,\rho )} \bigl[ \frac{1}{4} g_{\sigma}^2 (\rho )\Theta_{\sigma}
(k,q) - g_{\omega}^2(\rho )\Theta_{\omega}(k,q) \bigr] \label{eq:one}
\end{eqnarray}
using the notation of Serot and Walecka (Ref.~\cite{serot}, pp.~130-131)
The first line in this equation is the Hartree
contribution and the second line the Fock term. At each density, we
adjust the coupling constants $g_{\sigma}$ and $g_{\omega}$ such that
the scalar part of the self-energy and the total energy per
nucleon as obtained in the DBHF calculation for nuclear matter are
reproduced by a Hartree-Fock calculation using this simple
$\sigma$ - $\omega$ model. The resulting coupling constants
are reduced as compared to
the Hartree analysis, see Fig.~1. This can easily be understood
by comparing Eq.~(1) and the corresponding expression for the vector
component for the self-energy with the expressions
of the Hartree scheme. Notice that the density dependence is very similar
in both cases.

The density-dependent coupling constants deduced in the Hartree-Fock
analysis of the DBHF nuclear matter results can be employed in a
Dirac-Hartree-Fock calculation for finite nuclei. Comparing the
results of this relativistic, density-dependent Hartree-Fock (RDHF)
scheme with those obtained in the RDH approximation (see Table 1) one
observes that the Fock terms reduce the
binding energies and charge radii.
Thus, the very good agreement of the RDH
results with  experiment was just fortuitous and is lost when the
Fock terms are included.

There is, however,
good agreement between the
DBHF and RDHF results for $^{16}$O. For $^{40}$Ca, this agreement
appears less close.
However, one has to keep in mind that it
requires only a slight modification in the NN interaction to reduce the
energy and increase the radius, i.~e., to
``move'' the results parallel to the Coester band, whereas it is very
difficult to achieve a modification perpendicular to the Coester band.
In this sense, energies and radii calculated in DBHF and
RDHF are very close. This implies that both schemes, treating
either the Dirac effects or the effects of correlations in a local
density approximation, yield very similar results {\it if the Fock effects
are included}. This can be interpreted as an indication that DBHF
and RDHF are reliable approximations for a self-consistent
relativistic BHF calculation for energies and radii of finite nuclei.
In order to close the gap between the results obtained in DBHF (and
RDHF) and experiment it may be necessary to consider
three-nucleon correlations or other improvements of the many-body
approach beyond relativistic BHF.
Details in the spectrum of single-particle energies of RDHF should be
improved by taking additional mesons in the parameterization of the
effective NN interaction into account~\cite{elsen}.

In summary,
we have calculated the groundstate properties of $^{16}$O and $^{40}$Ca
using two different approximation methods for solving the relativistic
Brueckner-Hartree-Fock equations. The predictions, which are very similar
for both approximations,
are in substantially better agreement with experiment
as compared to conventional, non-relativistic
calculations. Fock terms must not be neglected.
Remaining small, but distinct discrepancies between theory and
experiment represent a challenge for future research in nuclear
many-body theory.

This work was supported in part by the Graduiertenkolleg T\"ubingen (DFG, Mu
705/3) and the US National Science
Foundation (PHY-9211607).

\vspace{16cm}
{\bf Fig.1:} { Density-dependent coupling constants deduced from the
Hartree and Hartree-Fock analysis of the DBHF results for nuclear
matter using the Bonn~A potential~[2].
The upper part of this figure displays
the meson-nucleon coupling constant for the scalar meson (with
$m_{\sigma}$=550 MeV), while the lower half shows the
result for the vector meson (with $m_{\omega}$= 783 MeV).}

\clearpage
{\bf Table 1.} Ground-state properties of $^{16}$O and
$^{40}$Ca.
The total energy per nucleon (E/A), the charge radius
($r_c$), and the
proton single-particle energies
($\epsilon_{i}$)
as predicted by non-relativistic
Brueckner-Hartree-Fock
(BHF), relativistic Dirac-BHF of Ref.~\cite{mmb} (DBHF),
relativistic, density-dependent Hartree (RDH), and relativistic,
density-dependent Hartree-Fock (RDHF) calculations are compared
to experiment (last column).
In all calculations the Bonn~A potential~\cite{rup} is used.

\begin{center}
\begin{tabular}{|c|rr|rr|r|}
\hline\hline
& \multicolumn{4}{c}{$^{16}$O} & \\
\hline
& BHF & DBHF & RDH & RDHF & Exp \\
\hline
&&&&&\\
E/A               [MeV] & -5.95 & -7.56 & -7.79 & -7.36 & -7.98 \\
r                 [fm]  & 2.31  & 2.46  &  2.67 &  2.47 & 2.70  \\
&&&&&\\
$\epsilon_{s1/2}$ [MeV] & -56.6 & -49.8 & -43.1 & -44.7 & -40$\pm$8 \\
$\epsilon_{p3/2}$ [MeV] & -25.7 & -23.0 & -22.3 & -23.8 & -18.4 \\
$\epsilon_{p1/2}$ [MeV] & -17.4 & -13.2 & -17.5 & -15.8 & -12.1 \\
&&&&&\\
\hline
& \multicolumn{4}{c}{$^{40}$Ca} & \\
\hline
& BHF & DBHF & RDH & RDHF & Exp \\
\hline
&&&&&\\
E/A               [MeV] & -8.29 & -8.64 & -8.20 & -7.93 & -8.5 \\
r                 [fm]  &  2.64 &  3.05 &  3.35 & 3.13  & 3.50 \\
&&&&&\\
$\epsilon_{d5/2}$ [MeV] & -30.2 & -21.9 & -19.8 & -21.2 & -14$\pm$2 \\
$\epsilon_{1s1/2}$[MeV] & -24.5 & -13.8 & -15.4 & -14.2 & -10$\pm$1 \\
$\epsilon_{d3/2}$ [MeV] & -16.5 & -10.2 & -14.5 & -13.2 & -7$\pm$1 \\
&&&&&\\
\hline\hline
\end{tabular}
\end{center}
\end{document}